\documentclass[showpacs,floatfix,superscriptaddress,showpacs,twocolumn,pra,aps]{revtex4}
\usepackage{amssymb}
\usepackage{longtable,graphicx,epsfig,dcolumn}

\begin{document}

\title{New interpretation of the atomic spectra and other quantum phenomena:
A mixed mechanism of classical $LC$ circuits and quantum wave-particle
duality}
\author{X. Q. Huang}
\email{xqhuang@nju.edu.cn}
\affiliation{Department of Physics and National Laboratory of Solid State
Microstructures, Nanjing University, Nanjing 210093, China}
\date{\today }

\begin{abstract}
\noindent We study the energy conversion laws of the macroscopic
harmonic $LC $ oscillator, the electromagnetic wave (photon) and
the hydrogen atom. As our analysis indicates that the energies of
these apparently different systems obey exactly the same energy
conversion law. Based on our results and the wave-particle duality
of electron, we find that the atom in fact is a natural
microscopic $LC$ oscillator. In the framework of classical
electromagnetic field theory we analytically obtain, for the
hydrogen atom, the quantized electron orbit radius
$r_{n}=a_{0}n^{2}$, and quantized energy $E_{n}=-R_{H}hc/n^{2},$
($n=1,2,3,\cdots ),$ where $a_{0}$ is the Bohr radius and $R_{H}$
is the Rydberg constant. Without the adaptation of any other
fundamental principles of quantum mechanics, we present a
reasonable explanation of the polarization of photon, the Zeeman
effect, Selection rules and Pauli exclusion principle.
Particularly, it is found that a pairing Pauli electron can move
closely and steadily in a DNA-like double helical electron orbit.
Our results also reveal an essential connection between electron
spin and the intrinsic helical movement of electron and indicate
that the spin itself is the effect of quantum confinement. In
addition, a possible physical mechanism of superconductivity and
the deeper physical understandings of the electron mass, zero
point energy (ZPE), and the hardness property of electron are also
provided. Finally, we show analytically that the Dirac's
quantization of magnetic monopole is merely a special handed
electron at absolute zero-temperature with the de Broglie
wavelength $\lambda _{e}=0.$ This is a new and surprising result,
which strongly suggests that any efforts to seek for the magnetic
monopole in real space will be entirely in vain. Furthermore, it
appears that the electron's spin and the magnetic monopole are
actually two different concepts for one possible physical
phenomenon.
\end{abstract}

\maketitle

\section{Introduction}

No one doubt that twentieth century is the century of quantum theory [1-13].
After 100 years of development quantum physics is no longer just a field, it
is the bedrock of all of modern physics. Although the modern quantum theory
has provided a beautiful and consistent theory for describing the myriad
baffling microphenomena which had previously defied explanation \cite{dirac}%
, one should not neglect a curious fact that quantum mechanics never take
into account the deep structures of atoms. In fact, at the heart of quantum
mechanics lies only the Schr\"{o}dinger equation \cite{schrodinger}, which
is the fundamental equation governing the electron. According to quantum
theory, it is the electromagnetic interaction (by the exchange of photons)
which hold electrons and nuclei together in the atoms. But, up to now,
quantum theory never provides a practical scheme that electron and nuclei
can absorb and emit photons.

In this paper, we investigate the energy relationship of electron in the
hydrogen atom. Significantly, we find a process of perfect transformation of
two forms of energy (kinetic and field energy) inside the atom and the
conservation of energy in the system. By applying the principle of
wave-particle duality and comparing to known results of the macroscopic
harmonic $LC$ oscillator and microscopic photon, we are assured that
electron kinetic energy in fact is a kind of magnetic energy and the atom is
a natural microscopic $LC$ oscillator. Moreover, the mixed mechanism
(classical $LC$ circuits / quantum wave-particle duality) turns out to have
remarkably rich and physical properties which can used to describe some
important quantum principles and phenomena, for instance, polarization of
photon, Zeeman effect, Selection rules, the electron's mass and spin, zero
point energy (ZPE) , the Pauli exclusion principle and the Dirac's magnetic
monopole.

\section{Energy transformation and conversion in hydrogen atom}

\begin{figure}[tbp]
\centering
\includegraphics[height=1.5in,angle=-90]{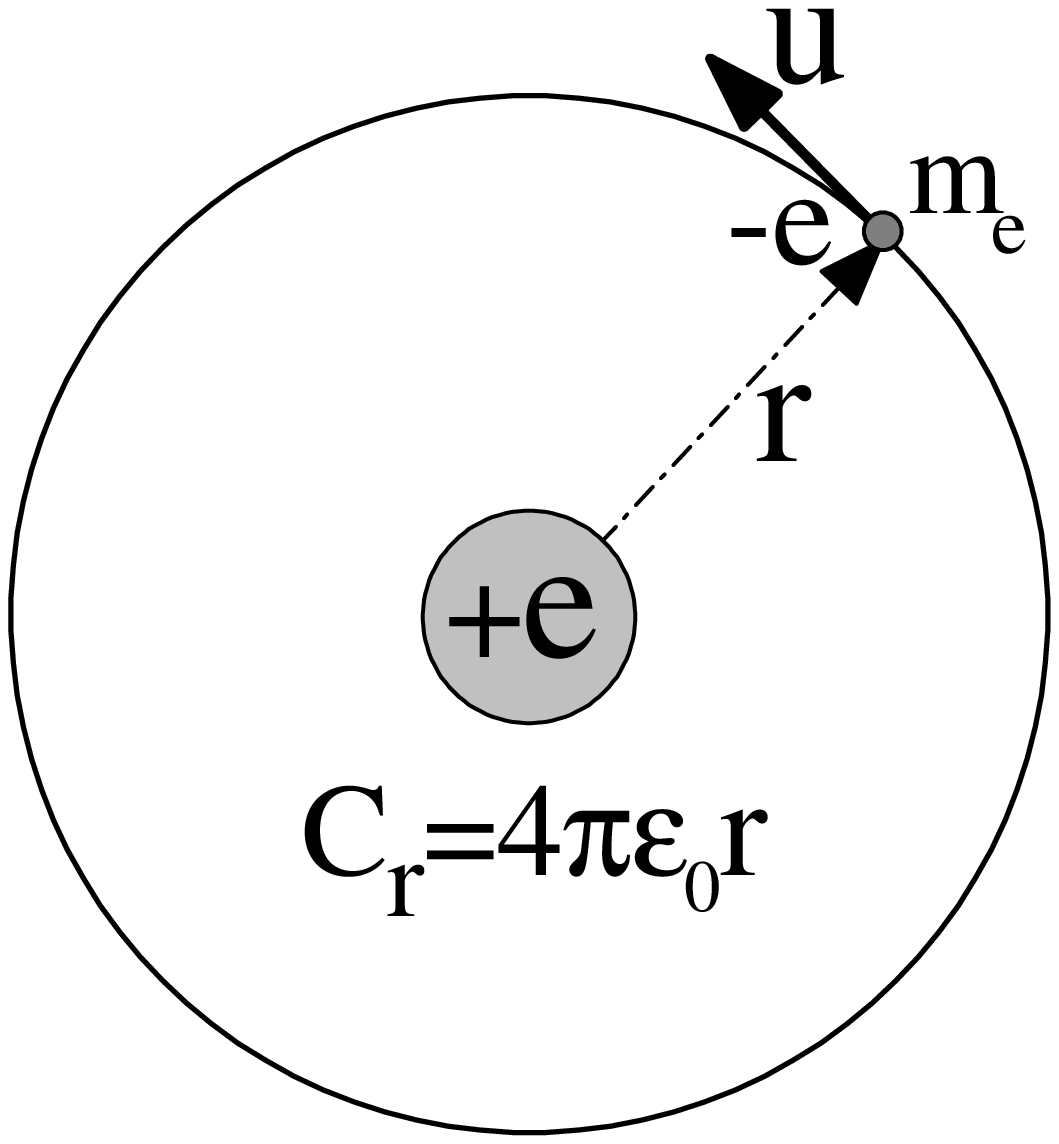}
\caption{The diagram illustrating the hydrogen atom}
\label{fig1}
\end{figure}
Classically, as shown in Fig. \ref{fig1}, the hydrogen atom consists of one
electron in orbit around one proton with the electron being held in place
via the electric Coulomb force. Equation of motion is
\begin{equation}
\frac{e^{2}}{4\pi \varepsilon _{0}r^{2}}=m_{e}\frac{u^{2}}{r},
\label{motion}
\end{equation}%
where $m_{e}$ is mass of electron. Eq. \ref{motion} can be rewritten in the
form of kinetic energy $E_{k}$ and field energy $E_{f}$ (stored in the
capacitor of hydrogen atom) as follows:
\begin{equation}
\frac{e^{2}}{2C_{r}}=\frac{1}{2}m_{e}u^{2},  \label{energy1}
\end{equation}%
where $C_{r}=4\pi \varepsilon _{0}r$ is the capacitance of the hydrogen
system. Thus the total energy of the hydrogen system is given by%
\begin{equation}
E_{total}=\left\vert \frac{1}{2}m_{e}u^{2}-\frac{e^{2}}{4\pi \varepsilon
_{0}r}\right\vert =\frac{e^{2}}{2C_{r}}.  \label{energy2}
\end{equation}

It should be pointed out that Eq. \ref{energy1} and \ref{energy2} are the
foundation of our study. These two equations together indicate a process of
perfect periodically transformation of two forms of energy (kinetic energy $%
E_{k}=\frac{1}{2}m_{e}u^{2}$ and field energy $E_{f}=e^{2}/2C_{r}$) inside
the atom and the conservation of energy in the system
\begin{equation}
E_{total}=E_{f}=E_{k}.  \label{conservation}
\end{equation}

Recall the macroscopic harmonic $LC$ oscillator where two forms of energy,
the maximum field energy $E_{f}=Q_{0}^{2}/2C$ of the capacitor $C$ (carrying
a charge $Q_{0}$) and the maximum magnetic energy $E_{m}=LI_{0}^{2}/2$ of
the inductor $L$, are mutually exactly interconvertible ($%
E_{total}=E_{f}=E_{m})$ with a exchange periodic $T=2\pi \sqrt{LC}$. And for
a microscopic photon (electromagnetic wave), the maximum field energy $E_{f}=%
\frac{1}{2}\varepsilon _{0}E_{0}^{2}$ and the maximum magnetic energy $E_{m}=%
\frac{1}{2}\mu _{0}H_{0}^{2}$ also satisfy $E_{total}=E_{f}=E_{m}$.

Based on the above energy relationship for three totally different systems
and the requirement of the electromagnetic interaction (by exchanging
photon) between electron and nuclei, we assure that the kinetic energy of
electron (Eq. \ref{energy1}) is a kind of magnetic energy and the hydrogen
atom is a natural microscopic $LC$ oscillator. In 2000, a multinational team
of physicists had observed for the first time a process of internal
conversion between bound atomic states when the binding energy of the
converted electron becomes larger than the nuclear transition energy \cite%
{carreyre,kishimoto}. This observation indicate that energy can pass
resonantly between the nuclear and electronic parts of the atom by a
resonant process similar to that which operates between an inductor and a
capacitor in an $LC$ circuit. These experimental results can be considered a
conclusive evidence of reliability of our $LC$ mechanism.

Here raise an important question: how can the electron function as an
excellent microscopic inductor? We think the answer lies in the intrinsic
wave-particle duality nature of electron. In our opinion, the wave-particle
nature \cite{broglie} of electron is only a macroscopic behavior of the
intrinsic helical motion of electron within its world.

\section{Chirality and \textquotedblleft inducton\textquotedblright\ of free
electron}

In 1923, Broglie suggested that all particles, not just photons, have both
wave and particle properties \cite{broglie}. The momentum wavelength
relationship for any material particles was given by

\begin{equation}
\lambda =\frac{h}{p}\mathbf{,}  \label{broglie}
\end{equation}%
where $\lambda $ is called de Broglie wavelength, $h$ is Planck's constant
\cite{planck} and $p$ the momentum of the particle. The subsequent
experiments established the wave nature of the electron \cite%
{davisson,thomson}. Eq.\ref{broglie} implies that, for a particle moving at
high speed, the momentum is large and the wavelength is small. In other
words, the faster a particle moves, the shorter is its wavelength.
Furthermore, it should be noted that any confinement of the studied particle
will shorten the $\lambda $ and help to enhance the so-called quantum
confinement effects.

\begin{figure}[tbp]
\centering
\includegraphics[height=3.4in,angle=-90]{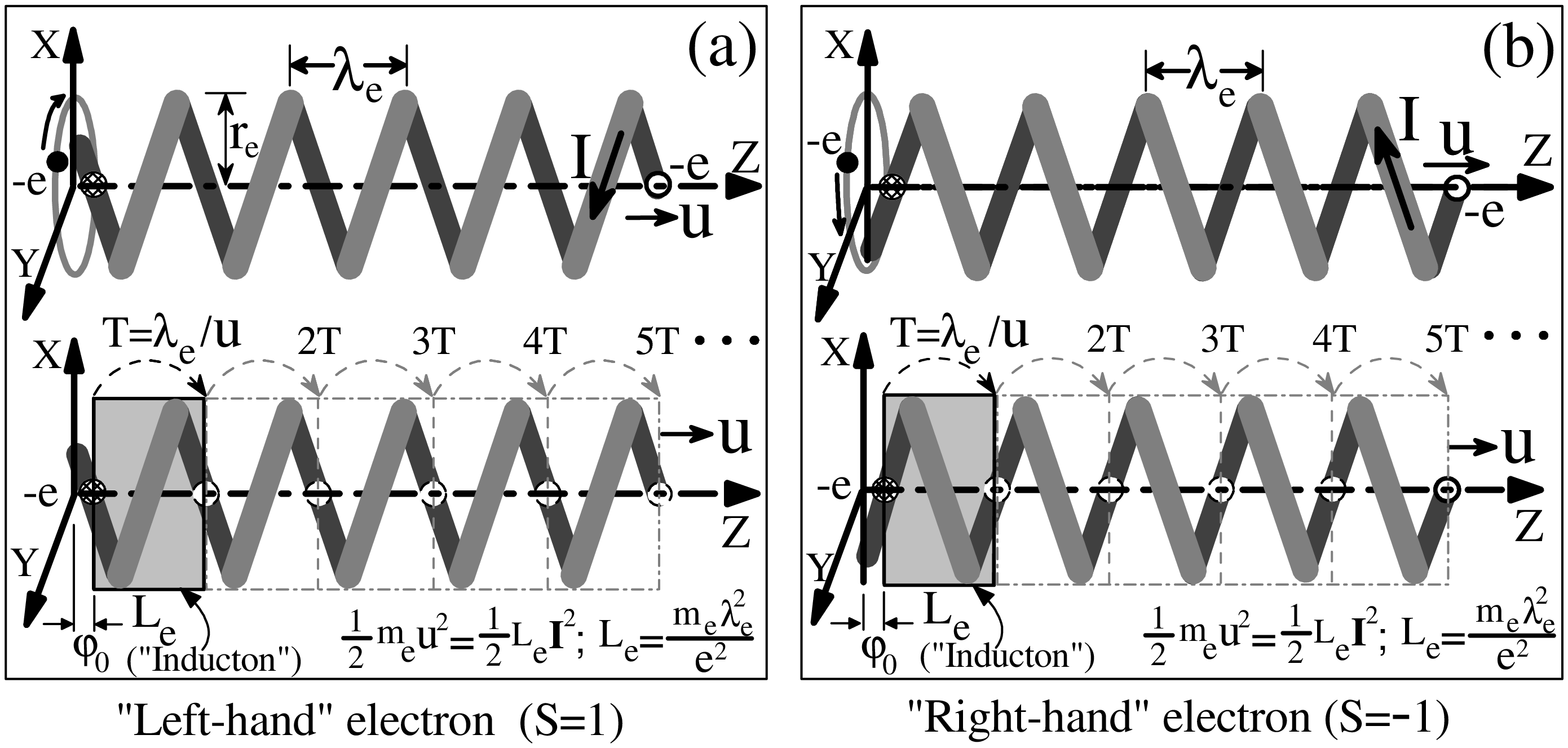}
\caption{A free electron moving along a helical orbit with a helical pitch
of de Broglie wavelength $\protect\lambda _{e}.$}
\label{fig2}
\end{figure}
As shown in Fig. \ref{fig2}(a) and (b), based on Eq. \ref{broglie} and the
demanding that the electron would be a microscopic inductor, we propose that
a free electron can move along a helical orbit (the helical pitch is de
Broglie wavelength $\lambda _{e})$ of left-handed or right-handed. In this
paper, the corresponding electrons are called \textquotedblleft
Left-hand\textquotedblright\ and \textquotedblleft
Right-hand\textquotedblright\ electron which are denoted by Chirality
Indexes $S=1$ and $S=-1,$ respectively. Hence, the electron can now be
considered as a periodic-motion quantized inductive particle which is called
\textquotedblleft inducton\textquotedblright\ (see Fig. \ref{fig2}).
Moreover, the particle-like kinetic energy of electron can be replaced with
a dual magnetic energy carried by a \textquotedblleft
inducton\textquotedblright . Therefore, we have
\begin{equation}
\mathbf{E}_{k}=\frac{1}{2}m_{e}u^{2}=\frac{1}{2}L_{e}\mathbf{I}^{2},
\label{energy0}
\end{equation}%
where $u$ is the axial velocity of the helical moving electron and $L_{e}$
is the inductance of the quantized \textquotedblleft
inducton\textquotedblright . The above relation indicates that the mass of
electron is associated with an amount of magnetic energy.

From Fig. \ref{fig2}, the electric current, for one de Broglie wavelength,
is given by
\begin{equation}
\mathbf{I}=\frac{eu}{\lambda _{e}}.  \label{current}
\end{equation}%
From Eq. \ref{current}, it is important to note that the electric current
should be defined within an integral number of de Broglie wavelength. Hence,
the electric current $\mathbf{I}=eu/2\pi r$ (where $r$ is the electronic
orbital radius in the hydrogen atom), which was widely used in the
semiclassical Bohr model, may be physically invalid. Collecting Eq. \ref%
{energy0} and \ref{current} together, we have the inductance of single
\textquotedblleft inducton\textquotedblright

\begin{equation}
L_{e}\mathbf{=}\frac{m_{e}\lambda _{e}^{2}}{e^{2}}.  \label{inducton}
\end{equation}%
Then the dual nature of electron can be uniquely determined by $L_{e},$ the
periodic $T$ (or frequency $f=1/T=u/\lambda _{e})$, the initial phase $%
\varphi _{0}$ and the chirality ($S=1$ or $S=-1$).

\section{Atomic spectra of hydrogen atom}

\subsection{Quantized radius and energy}

By the application of helical electron orbit to the hydrogen atom (Fig. \ref%
{fig2}), we can not only explain the stability of the atom but also give a
theoretical interpretation of the atomic spectra.

\begin{figure}[tbp]
\centering
\includegraphics[height=3.2in,angle=-90]{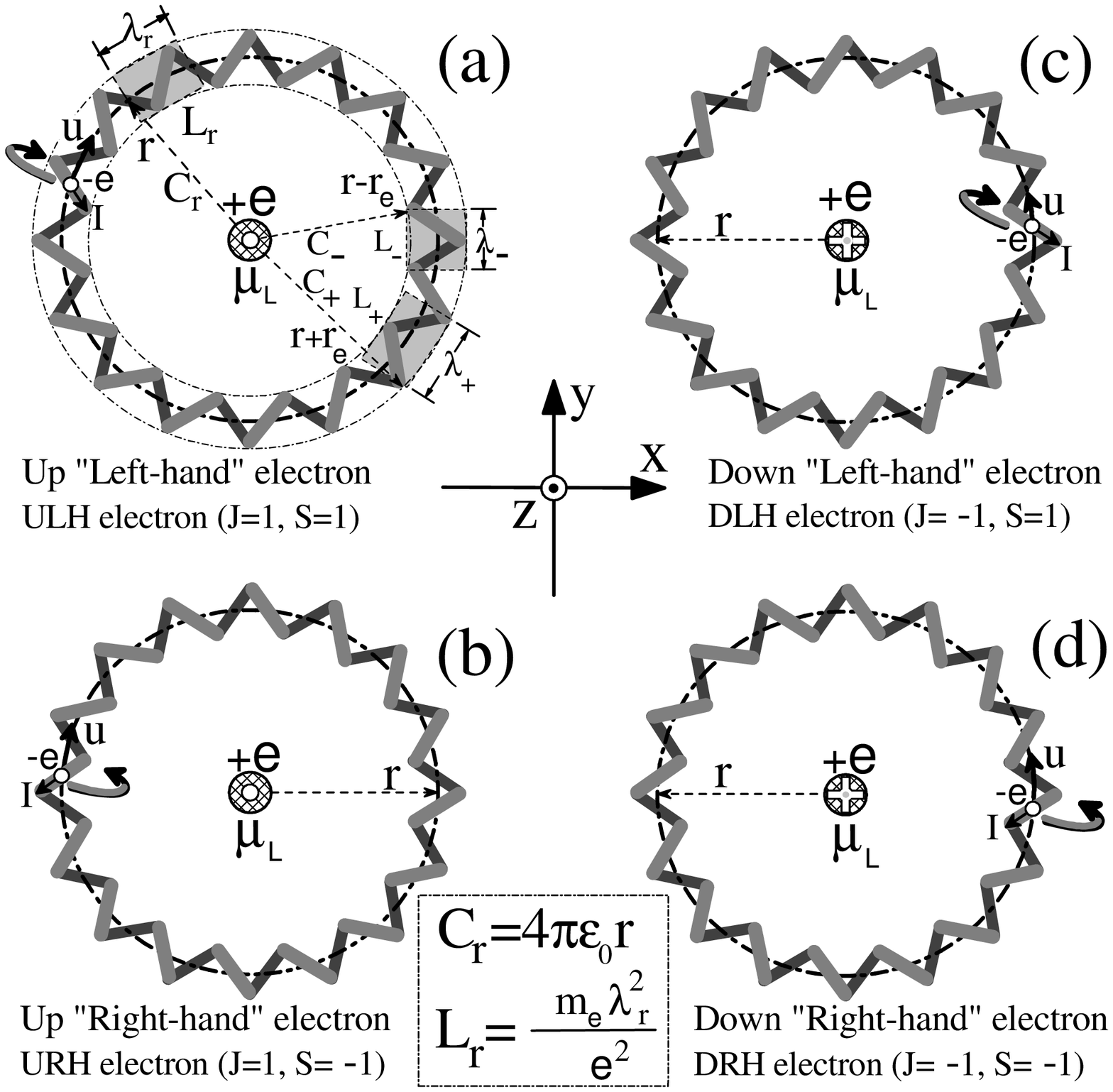}
\caption{The quadruple degenerate stable helical electron orbits in hydrogen
atom.}
\label{fig3}
\end{figure}

Fig. \ref{fig3} shows four possible kinds of stable helical electron orbits
in hydrogen atom, and each subgraph corresponds to a electron of different
motion manner within the atom. The electrons can be distinguished by the
following two aspects. First consider the chirality of electron orbits, as
shown in Fig. \ref{fig3}, the electrons of Fig. \ref{fig3}(a) and (c) are
\textquotedblleft Left-hand\textquotedblright\ labelled by $S=1$, while
electrons of Fig. \ref{fig3}(b) and (d) are \textquotedblleft
Right-hand\textquotedblright\ labelled by $S=-1$. Secondly consider the
direction of electron orbital magnetic moment $\mu _{L}$, Fig. \ref{fig3}(a)
and (b) show that the $\mu _{L}$ are in the $Z$ direction (Up) while (c) and
(d) in the $-Z$ direction (Down), the corresponding electrons are labelled
by $J=1$ and $J=-1,$ respectively, here $J$ is called Magnetic Index. Hence,
the electrons of different physical properties become distinguishable, they
are Up \textquotedblleft Left-hand\textquotedblright\ (ULH) electron ($J=1$,
$S=1$), Up \textquotedblleft Right-hand\textquotedblright\ (URH) electron ($%
J=1$, $S=-1$), Down \textquotedblleft Left-hand\textquotedblright\ (DLH)
electron ($J=-1$, $S=1$) and Down \textquotedblleft
Right-hand\textquotedblright\ (DRH) electron ($J=-1$, $S=-1$).

As shown in Fig. \ref{fig3}(a), the helical moving electron around
the orbit mean radius $r$ can now be regarded as a quantized
\textquotedblleft inducton\textquotedblright\ of $\lambda _{r},$
thus the hydrogen atom is a natural microscopic $LC$ oscillator.
We consider that the physical
properties of the hydrogen atom can be uniquely determined by these natural $%
LC$ parameters. To prove that our theory is valid in explaining
the structure of atomic spectra, we study the quantized orbit
radius and the quantized energy of hydrogen atom and make a
comparison between our results of $LC$ mechanism and the known
results of quantum theory. For the system of $\lambda _{r}$, the
$LC$ parameters of the hydrogen atom is illustrated in Fig.
\ref{fig3}. Then the $LC$ resonant frequency is

\begin{equation}
\nu _{r}=\frac{1}{2\pi \sqrt{L_{r}C_{r}}}=\frac{e}{4\pi \lambda _{r}\sqrt{%
\pi \varepsilon _{0}m_{e}r}}.  \label{frequency}
\end{equation}%
Recall the well-known relationship $E=h\nu _{r},$ we have
\begin{equation}
h\nu _{r}=\frac{e^{2}}{8\pi \varepsilon _{0}r}.  \label{energy3}
\end{equation}%
Combining Eq. \ref{frequency} and Eq. \ref{energy3} gives
\begin{equation}
\lambda _{r}=\frac{2h}{e}\sqrt{\frac{\pi \varepsilon _{0}r}{m_{e}}}.
\label{wavelength}
\end{equation}%
Then the stable electron orbits are determined by
\begin{equation}
\frac{2\pi r}{\lambda _{r}}=n,(n=1,2,3\cdots ),  \label{stable}
\end{equation}%
where $n$ is called Principal oscillator number. The integer $n$
shows that the orbital allow integer number of \textquotedblleft
induction\textquotedblright\ of the de Broglie wavelength $\lambda
_{r}$. From Eq. \ref{wavelength} and Eq. \ref{stable}, the
quantized electron orbit mean radius is given by

\begin{equation}
r_{n}=\frac{\varepsilon _{0}h^{2}}{\pi m_{e}e^{2}}n^{2}=a_{0}n^{2},
\label{radius}
\end{equation}%
where $a_{0}$ is the Bohr radius. And the quantized energy is

\begin{equation}
E_{n}=-\frac{e^{2}}{8\pi \varepsilon _{o}r_{n}}=-\frac{m_{e}e^{4}}{%
8\varepsilon _{0}^{2}h^{2}}\frac{1}{n^{2}}=-R_{H}\frac{hc}{n^{2}},
\label{energy5}
\end{equation}%
where $R_{H}$ is the Rydberg constant. Surprisingly, the results of Eq. \ref%
{radius} and \ref{energy5} are in excellent agreement with Bohr model \cite%
{bohr}. Besides, taking Fig. \ref{fig3} into account, we can conclude that
the quantized energies of Eq. \ref{energy5} are quadruple degenerate.

\subsection{Lamb shift and polarization of photon}

Now, in the framework of helical electron orbit (see Fig. \ref{fig3}), the
electron orbital magnetic moment $\mu _{L}$ is allowed naturally to be both
positive ($J=1$) and negative ($J=-1$), therefore, the double splitting
experiment \cite{stern} is a immediate result of our theory. It seems that
the electron spin \cite{uhlenbeck} no long as an essential quantum number in
our studies. In Section \ref{discus}, we will try to explain for the first
time the connection between the electron spin and the intrinsic helical
movement of electron.

Furthermore, take a look at Fig. \ref{fig3}(a), the other two
\textquotedblleft inductons\textquotedblright\ with different de Broglie
wavelengths $\lambda _{-}$, and $\lambda _{+}$ are defined. Here we should
stress that $\lambda _{r}$ in fact is invalid in a system of circular
motion. To illustrate this ideas, let us examine the three situations ($%
\lambda _{-}$, $\lambda _{r}$ and $\lambda _{+}$) carefully. For the case of
$\lambda _{r},$ both ends of the \textquotedblleft
inducton\textquotedblright\ doesn't fall in the $XY$ plane and electron
travelling through these two locations has a nonzero radial velocity, in
this case we think the quantized \textquotedblleft
inducton\textquotedblright\ defined by $\lambda _{r}$ is unstable thus
invalid in the system. But, for the other two cases of $\lambda _{+}$ and $%
\lambda _{-}$, both ends of these \textquotedblleft
inductons\textquotedblright\ fall in the $XY$ plane, at the same time the
corresponding radial velocity of the travelling electrons is zero, in other
words, these two \textquotedblleft inductions\textquotedblright\ described
by $\lambda _{+}$ and $\lambda _{-}$ are stable and can function as the
\textquotedblleft inductions\textquotedblright\ of the studied system. Hence
for a given quantized electron orbit mean radius of Eq. \ref{radius}, the
microscopic atom can contribute two natural $LC$ oscillators ($L_{+}C_{+}$
and $L_{-}C_{-}$ of Fig. \ref{fig3}(a)). In our opinion, it is these two
oscillators\ of $\lambda _{+}$ and $\lambda _{-}$ that finally lead to the
Lamb shift \cite{lamb} and the Bohr theory of hydrogen electron orbits ($%
\lambda _{r})$ are merely a approximate treatment of the corresponding
helical electronic orbits.

The polarization of photon perhaps still is the greatest mystery in the
microscopic world. When we reduce the light intensity to its smallest
possible level, then we are dealing with one single photon $-$ the quantized
light. Now a great number of the quantized light test experiments have show
that, in reality, the intrinsic polarization of photon is purely circular:
either right or left circular polarization. Quantum mechanics predicts that
the \textquotedblleft polarization\textquotedblright\ is related to the
electronic spin. Nevertheless, this mystical explanation is somewhere far
beyond the reality of the physical world. We state that the polarization of
quantized photon is referred to the intrinsic helical movement of electron
in the atom (Fig. \ref{fig3}). Namely, the left-hand electron ($S=1$) will
emit only left circular polarized photon, while right-hand electron ($S=-1$)
will emit right circular polarized photon.

\subsection{ \emph{LC} oscillator and quantum numbers}

\begin{figure}[bp]
\centering
\includegraphics[height=3.2in,angle=-90]{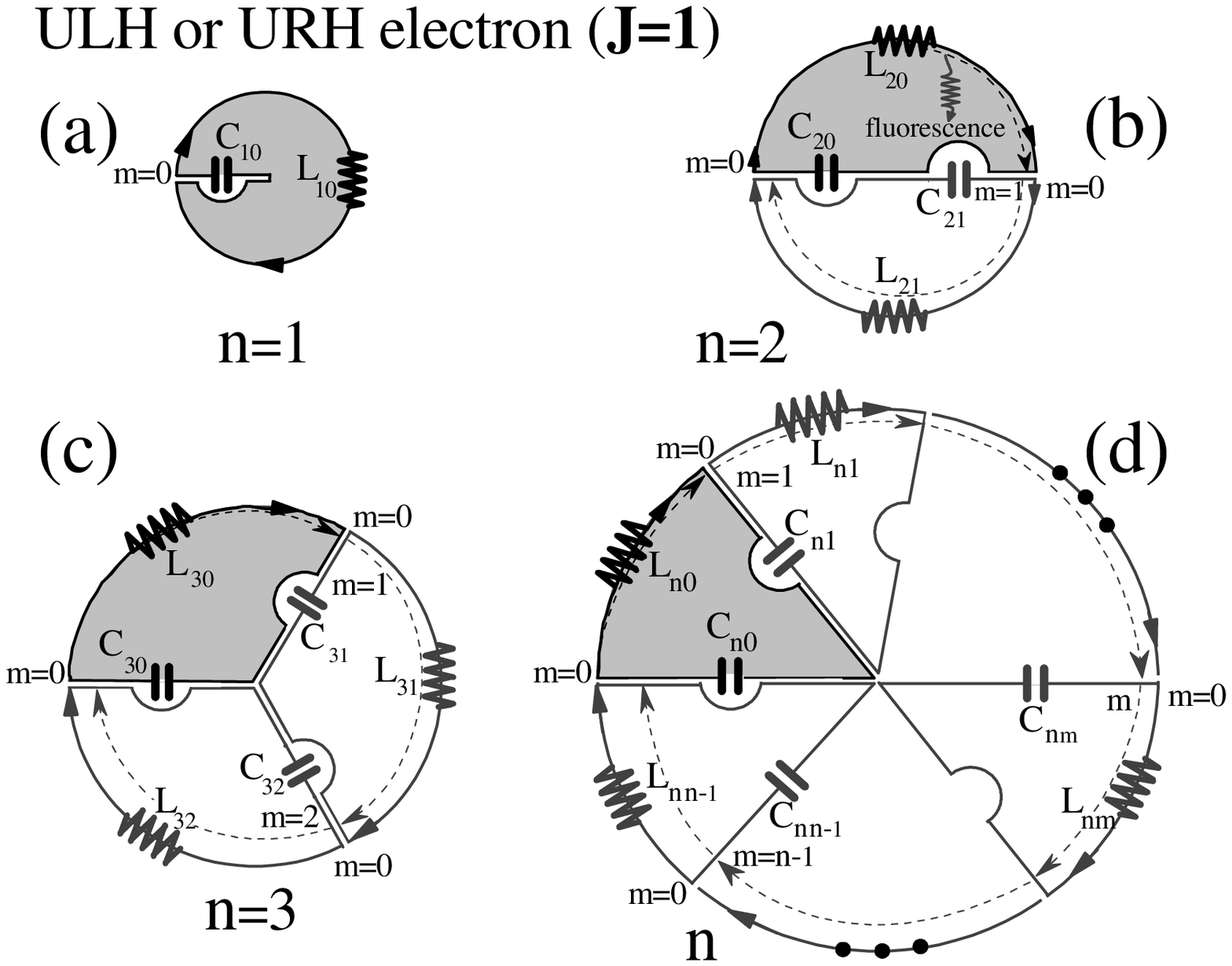}
\caption{The electron travelling orbits represented by the LC oscillators. }
\label{fig4}
\end{figure}

In quantum theory, four quantum numbers (the principal quantum number $n$,
the angular momentum quantum number $l$, the magnetic quantum number $m$ and
the spin quantum number $s)$ form a complete set used to specify the full
quantum state of any system in quantum mechanics. To verify that our $LC$
mechanism is designed to completely replace the quantum mechanism, a similar
perfect set of indexes (or quantum numbers) must be physically defined.
Recall the above discussions, three indexes (the Chirality Index $S$, the
Magnetic Index $J$ and the Principal Oscillator Number $n$) have been
introduced. Our three indexes (or quantum numbers) are not a complete set $-$
that is, they are insufficient to fully specify the quantum state of the
atom.
\begin{table*}[tp]
\caption{A comparison of the quantum numbers, eigenenergies of the quantum
mechanism and the quantum indexes, resonant frequencies of the LC mechanism.
}
\label{table1}\centering%
\begin{tabular}{cc|cc}
\hline
\multicolumn{2}{c|}{Quantum Mechanism} & \multicolumn{2}{|c}{$LC$ Mechanism}
\\ \hline
Quantum numbers & Eigenenergies & Quantum indexes & Resonant frequencies \\
\hline
$n$ & \multicolumn{1}{l|}{$E_{n}$, $n=1,2,3\cdots $} & $n$ &
\multicolumn{1}{l}{$\nu _{n},n=1,2,3...$} \\
$l$ & \multicolumn{1}{l|}{$E_{nl}$, $l=0,1,2,\cdots n-1$} & $l$ &
\multicolumn{1}{l}{$\nu _{nl},l=0,1,2,\cdots n-1$} \\
$m$ & \multicolumn{1}{l|}{$E_{nl}(m),m=0,\pm 1,\cdots \pm l$} & $m,J$ &
\multicolumn{1}{l}{$\nu _{nl}(m),m=0,\pm 1,\cdots \pm l$} \\
&  &  & $(m\geq 0$ for $J=1,$while $m\leq 0$ for $J=-1)$ \\
$s$ & \multicolumn{1}{l|}{$E_{nl}(m,s)$ $s=\frac{1}{2},-\frac{1}{2}$} & $S$
& \multicolumn{1}{l}{$\nu _{nl}(m,S),S=1,-1$} \\ \hline
\end{tabular}%
\end{table*}

Our goal here is to make the quantum numbers of the $LC$ mechanism towards
integrality. As shown in Fig. \ref{fig4}, we represent the hydrogen atom of
different states in the form of $LC$ oscillators. Fig. \ref{fig4}(a)$-$(d)
are the results for $n=1,$ $2,$ $3$ and $n$ of the ULH or URH electron ($J=1$
of Fig. \ref{fig3}$)$, respectively. In each subgraph, the gray oscillator
is the initial one and the solid line indicates an ideal radiationless
moving oscillator. In this stable situation, we have $C_{nl}=C_{n0}$ and $%
L_{nl}=L_{n0}$ (where $l=0,1,2,\cdots n-1)$ and the corresponding atom can
be regarded as a $n-fold$ degenerate $LC$ oscillator, thus the resonant
frequency as defined in Eq. \ref{frequency} is also $n-fold$ degenerate.
When interfered by extraneous factor, the electron will lose a small amount
of energy by emitting fluorescence and departure gradually from the original
circular motion and get closer to nuclei (shown by the dash lines of Fig. %
\ref{fig4}), consequently, we have $C_{n,n-1}<\cdots C_{n1}<C_{n0}$ and $%
L_{n,n-1}<\cdots L_{n1}<L_{n0}$ thus the original $n-fold$ degenerate $LC$
oscillator will split into $n$ non-degenerate oscillators $L_{nl}C_{nl}$,
where $l=0,1,2,\cdots n-1,$ which is called Metastable Oscillator Number.
For the DLH or DRH electron ($J=-1$)$,$ similarly we can also obtain $n$
non-degenerate $LC$ oscillators. Let us combine these results ($J=1$ and $%
J=-1$) thus for a given $n$ of Eq. \ref{stable}, there are at most $2n-1$
metastable $LC$ oscillators which are labelled uniquely by $m=0,1,2,\cdots
,l,$ and $m=0,-1,-2,\cdots ,-l$ for $J=1$ and $J=-1$, respectively. As a
result, we claim that the concept of the so-called magnetic quantum number
of quantum theory actually is the number of metastable $LC$ oscillators
existing in the atom. Now we include all quantum numbers required by the $LC$
mechanism. To show this, we summarize the above discussions and make a
comparison between the quantum numbers of the quantum mechanism and the
quantum indexes of the $LC$ mechanism. As shown in Table. \ref{table1}, the
resonant frequencies (energies) of the electron in the atom are uniquely
defined by $\nu _{nl}(m,S).$Without the external interference, $\nu
_{nl}(m,S)$ ($J=1,-1$ and $S=1,-1$) will keep its quadruple degenerate
characteristic.

\subsection{Zeeman effect}

\begin{figure}[tbp]
\centering
\includegraphics[height=3.0in,angle=-90]{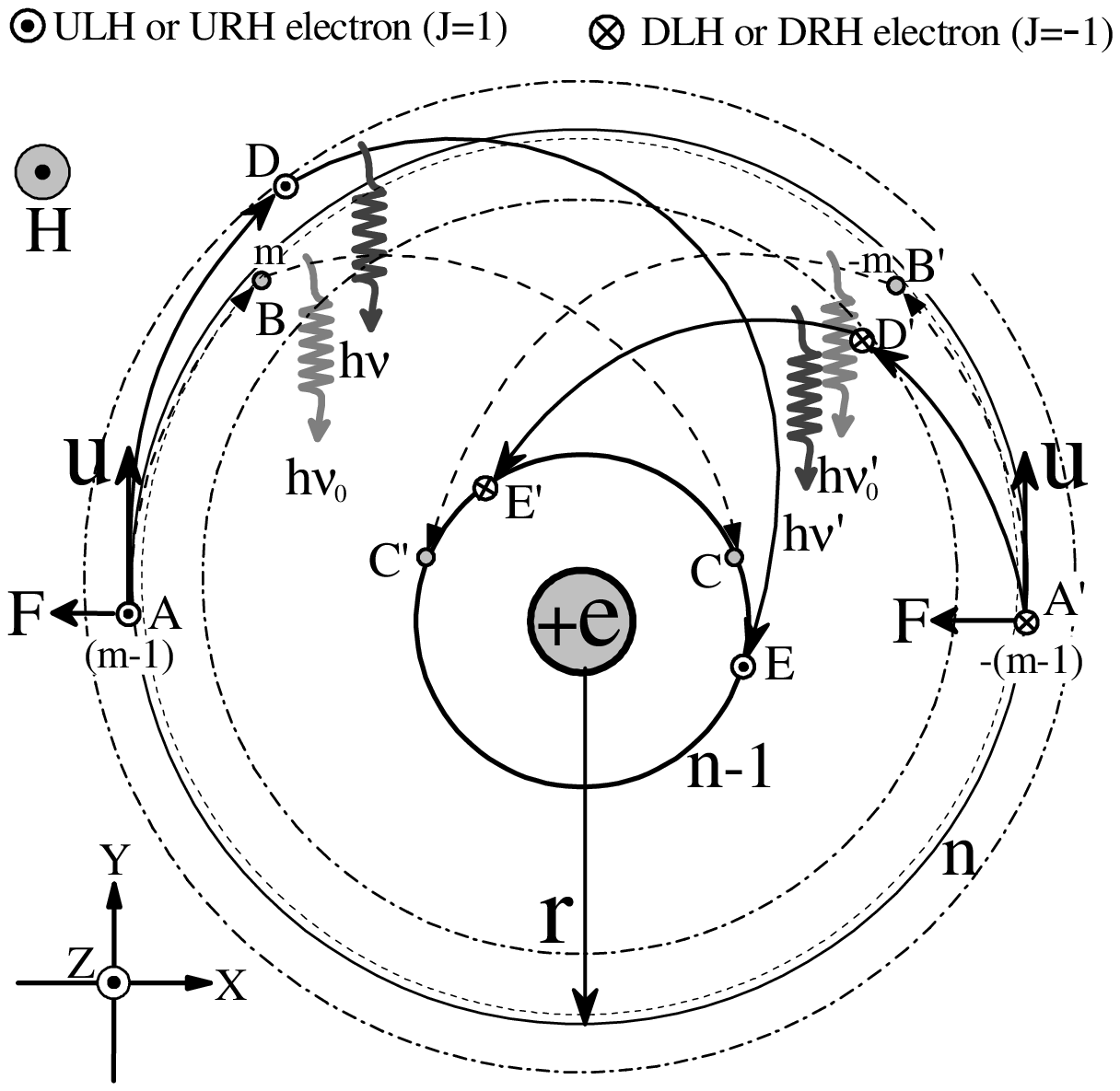}
\caption{A schematic interpretation of the Zeeman effect }
\label{fig5}
\end{figure}
When atomic spectral lines are split by the application of an external
magnetic field, it is called the Zeeman effect \cite{zeeman}. The Zeeman
effect in hydrogen atoms showed the expected equally-spaced triplet. Quantum
theory explains this by the spin-orbit interaction. By Fig. \ref{fig5}, we
can also interpret qualitatively this puzzling experimental result. As shown
in Fig. \ref{fig5}, the initial helical electron (with the same quantum
indexes $n$ and $l$) can be in a metastable state: either $A$ of $m-1$ ($%
J=1) $ or $A\prime $ of $-(m-1)$ ($J=-1)$. Without the external magnetic
field, the electrons are likely decay to the next metastable state ($m$ or $%
-m$) by emitting fluorescence, then transfer to the lower electronic orbit ($%
n-1$). During the transition ($B\rightarrow C)$ or ($B^{\prime }\rightarrow
C^{\prime }),$ the electron will emit a photon of frequency $\nu _{0}$ or $%
\nu _{0}^{^{\prime }}$, where $\nu _{0}=\nu _{0}^{^{\prime }}$. When applied
an external magnetic field in $Z$ direction, the electron ($J=1$ or $m>0$)
will be pushed to a higher energy states ($A\rightarrow D)$ by the lorentz
force (centrifugal force) then the transition of the electron ($D\rightarrow
E)$ is accompanied by the emission of a photon of frequency $\nu >\nu _{0}$,
while the electron ($J=-1$ or $m<0$ ) will be pulled to the lower states ($%
A^{\prime }\rightarrow D^{\prime })$ and the emitting photon frequency (from
$D^{\prime }$ to $E^{\prime }$) is $\nu ^{^{\prime }}<\nu _{0},$
consequently, the original single line ($\nu _{0})$ will be triplet ($\nu
^{^{\prime }},\nu _{0},\nu )$. The lines corresponding to Zeeman splitting
also exhibit polarization effects. According to our $LC$ mechanism, the
basic physical reason of the polarization is the intrinsic helical movement
of electron. Without magnetic field, the line ($\nu _{0})$ is a mix of equal
amounts of both right and left circular polarization photons, hence the line
($\nu _{0})$ is a nature light. When a magnetic field is applied (as shown
in Fig. \ref{fig5}), the side lines ($\nu ^{^{\prime }}$ and $\nu )$ are
circular polarization. This turns to be in good agreement with experimental
observations. A more detailed study of the Zeeman effect will involve the
selection rules which will be given in the following discussions.

\subsection{Selection rules}

\begin{figure}[tbp]
\centering
\includegraphics[height=3.0in,angle=-90]{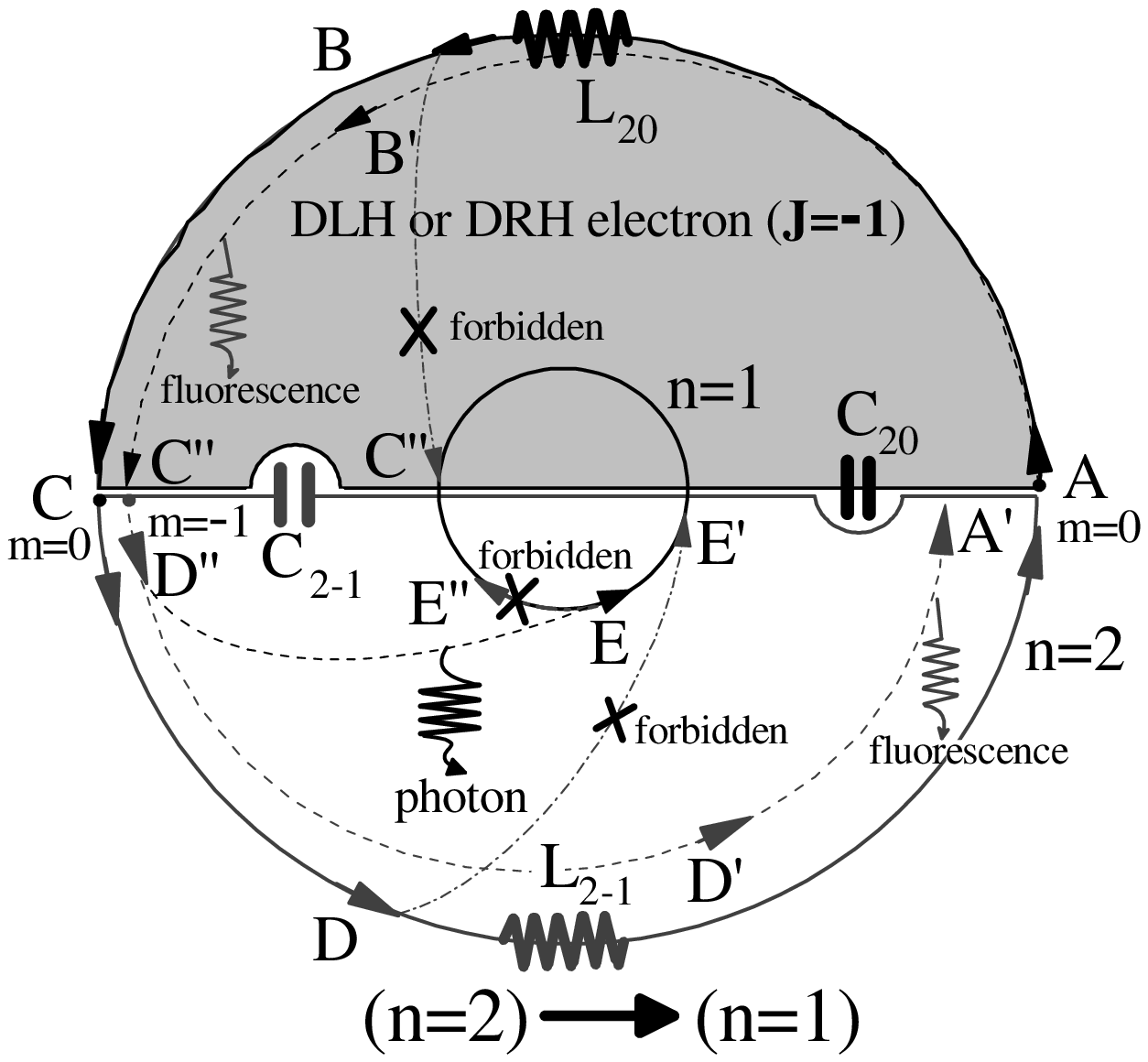}
\caption{A schematic interpretation of the Selection rules }
\label{fig6}
\end{figure}

In spectral phenomena, it becomes evident that transitions are not observed
between all pairs of energy levels. Some transitions are "forbidden" while
others are "allowed\textquotedblright\ by a set of selection rules. There
are three main selection rules for photon emission given by quantum theory.
These are $\Delta m=0,\pm 1,$ $\Delta s=0$ and $\Delta l=\pm 1.$What are the
selection rules under the $LC$ mechanism? Fig. \ref{fig6}. shows the
simplest transition ($n=1\rightarrow n=2)$ in the hydrogen atom. For the
electron of higher energy ($n=2$), there are two stable $LC$ oscillators (or
one double degenerate oscillator) for every circle, for convenience the
oscillators are described in another manner: $A\rightarrow B\rightarrow C$
and $C\rightarrow D\rightarrow A$ (see Fig. \ref{fig6}), respectively. As
can be seen, the direct transition of electron from a stable orbit, for
example $A\rightarrow B\rightarrow C^{\prime }$ and $C\rightarrow
D\rightarrow E^{\prime }$ are \textquotedblleft forbidden\textquotedblright
. When emitting fluorescence, the electron will decay to a slightly lower
orbit (shown by $A\rightarrow B^{\prime }\rightarrow C^{\prime \prime }$)
then the electron has a possibility of transition ($C^{\prime \prime
}\rightarrow D^{\prime \prime }\rightarrow E)$ to orbit of $n=1$ and emits a
photon, note that $\Delta J=0.$ While the following transition $C^{\prime
\prime }\rightarrow D^{\prime \prime }\rightarrow E^{\prime \prime }$ ($%
\Delta J\neq 0$) is \textquotedblleft forbidden\textquotedblright . We
cannot have a transition from $J=1$ to $J=-1 $, and vice versa, since the
transition involves a flip of the electron orbital magnetic moment which
must involve a change of flip energy. Another transition (unshown in the
figure) $S=1$ to $S=-1$ (and $S=-1$ to $S=1$) can't happen because there is
also a barrier energy which forbids the transition. Though we consider only
the simplest transition in the hydrogen atom, the forbidden transitions
described here are valid in other complex atom systems.

Quantum theory describes the selection rules by a too complicated schemes
which include spin-orbit coupling, wavefunction, transition probability and
total angular momentum quantum number, etc. Here, we only wanted to point
out that fundamental law of nature cannot be so complicated. The selection
rules expressed in quantum terms $\Delta m=0,\pm 1$ and $\Delta l=\pm 1$ are
only some characteristic parameters of the dominated and simplest electronic
transfer paths determined by the least-action principle. In other words, the
electron is likely to have a transition to the nearest states in a simplest
manner.

\section{Some discussions}

\label{discus}

To interpret atomic spectra completely and neatly, the Pauli exclusion
principle should be considered. In the following, firstly to describe the
Pauli exclusion principle in our theory, and secondly to provide a possible
physical mechanism of superconductivity and a deeper physical understanding
of the electron spin, electron mass, zero point energy (ZPE), the hardness
property of electron and magnetic monopole.

\subsection{Pauli exclusion principle}

\begin{figure}[tbp]
\centering
\includegraphics[height=3.2in,angle=-90]{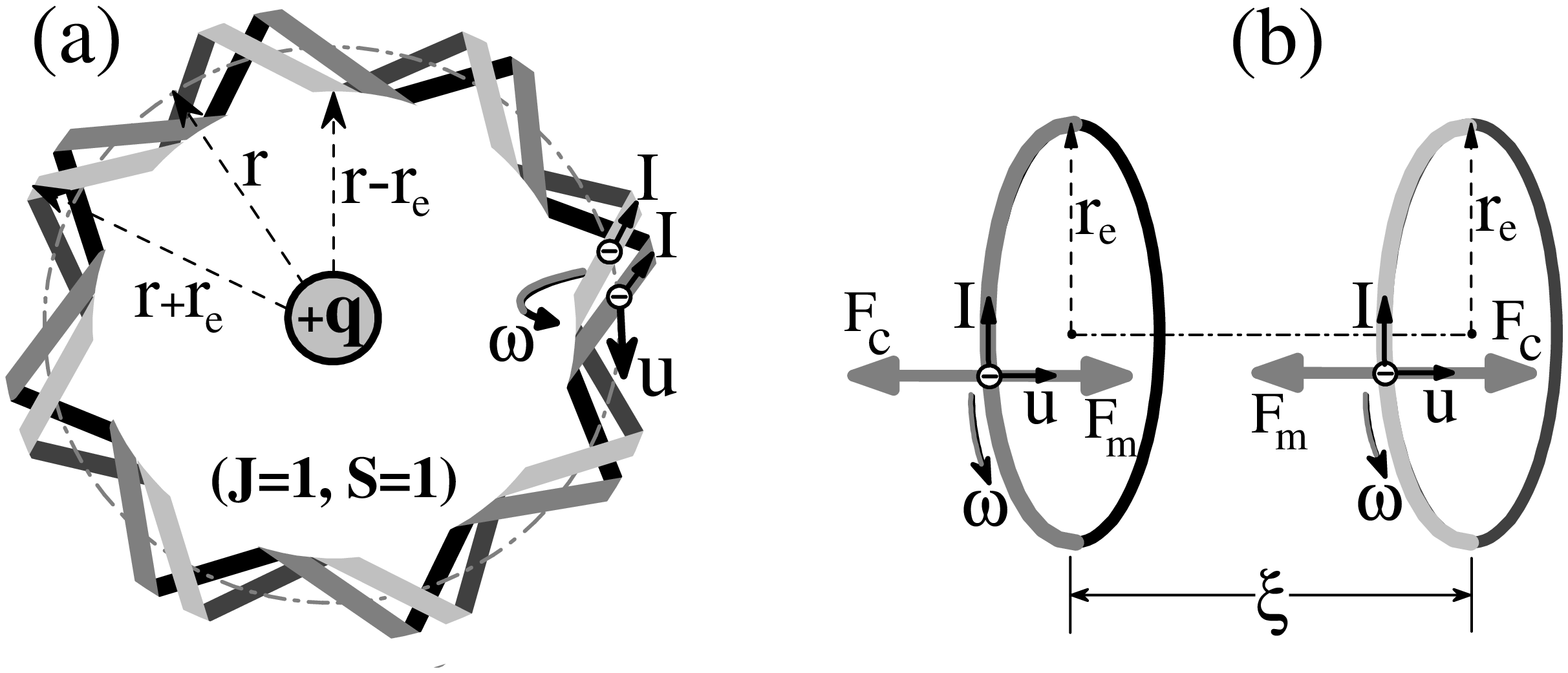}
\caption{The diagram illustrating the Pauli exclusion principle. (a) A
DNA-like double helical electron orbit, (b) a pairing Pauli electron.}
\label{fig7}
\end{figure}

Up to now, the Pauli exclusion principle is still puzzling to researchers.
It states that no two electrons in a single atom can have the same set of
quantum numbers. More significantly, Pauli's Exclusion Principle is not
enforced by any physical force understood by mainstream science.
\textquotedblleft When an electron enters an ion, somehow it knows the
quantum numbers of the electrons which are there, and somehow it knows which
atomic orbitals it may enter, and which not.\textquotedblright\ This is
nothing short of incredible! It implies consciousness or connectedness
between any and all elementary particles, and by a method totally unknown to
the mainstream purveyors of quantum physics. Our viewpoint about this
apparently esoteric principle is quite different, we think that the nature
of the Pauli exclusion principle can be illustrated by classical
electromagnetic field theory.

How do the Pauli electrons (pairing electrons) maintain stable orbits around
protons? According to the classical theory, if one electron is closer to
another electron, each electron will experience a repulsive electromagnetic
force which implies that it is impossible for these electrons to form a
stable Pauli electrons. But, the Pauli Exclusion Principle was shown
experimentally to be a more valid concept. Hence, this fact force us to seek
a attractive force between the Pauli electrons. As shown in Fig. \ref{fig7}%
(a), two electrons (with the same initial phase$)$ of the same chirality ($%
S=1$ or $S=-1$) can move closely in a DNA-like \cite{watson} double helical
electron orbit. Note that the two helical moving electrons are equivalent to
a pair of parallel electric currents, by the classical theory, these two
currents (or electrons) are attracted each other and the corresponding
electrons are called coherent electrons (or superconducting electrons). More
recently, Morris \textit{et al}. reported a astonishing observation of an
infrared nebula having the morphology of an intertwined double helix about
100 parsecs \cite{morris}. \ To our way of thinking, nature seems indeed
miraculous but really knowable. Some nature's laws and phenomena (for
example, the double helical structure) would be analogous at all scales,
microscopic, macroscopic or astroscopic scale. To the best of our knowledge,
it is for the first time that a possible double helical structure of
microworld is suggested and reported here.

To illuminate the physical scheme of the Pauli Exclusion Principle more
clearly in the classical theory, we try to simplify the picture of the
pairing electrons in the handed electron mechanism. As shown in Fig. \ref%
{fig7}(b), two same-phase helical moving electrons are also equivalent to a
pair of parallel travelling current ring. It is known that, for the two
static electrons, there is only a strong repulsive Coulomb force ($F_{c})$
between them, while for the two moving electrons, the new interaction will
be the magnetic force which can be attractive and repulsive depending on the
kinematic relation of the moving electrons. For the special case of Fig. \ref%
{fig7}, the magnetic forces $F_{m}$ exerted on the electrons are attractive,
and normally the two forces satisfy $F_{c}\gg F_{m} $. In order to form a
stable Pauli pairing electron, the condition $F_{c}=F_{m}$ should be
required. Therefore, the complicated question about the physical original
and the stability of the Pauli electrons may turn out to be a very simple
question: how to increase the magnetic force $F_{m}$ between the two helical
moving electrons? It has been known that $F_{m}\varpropto\mathbf{I}^{2}$,
thus, to demonstrate the reliability of our theory, the way to increase the
electric current $\mathbf{I}$ should be physically provided. From Eq. \ref%
{current}, we know that the confinement of electrons will shorten the $%
\lambda _{e}$ and hence greatly increase the electric current $\mathbf{I,}$
which eventually increase the attractive interaction between the two
same-phase helical moving electrons. As shown in Fig. \ref{fig7}(b), when
the two electrons are a certain distance apart ($\xi ,$ is called the
coherent length$)$, it is likely that the repulsive force ($F_{c})$ would be
equal to the attractive force ($F_{m}$). If more electrons are added in the
same helical orbit, thus these new injected electrons will subject a strong
repulsive Coulomb force from the existed Pauli electrons. The repulsive
Coulomb force can force these new added electrons to depart from the
occupied Pauli orbit. From our point of view, this fundamental
characteristic of the helical electrons is the physical nature of Pauli
Exclusion Principle.

\subsection{Superconductivity}

Now, we would like to give a brief discussion about superconductivity. It is
no doubt that the studies of the mechanism of superconductivity were, and
still are, a challenging physical problem. As shown in Fig. \ref{fig7}, our
mechanism provides a vivid and solid physical picture where two electrons
can be in pairing. We assume, for any kinds of superconductors, that the $%
T_{c}$ has the form: $T_{c}\thicksim \lambda _{e}^{-\eta }$ , where $\lambda
_{e}$ is the electron's de Broglie wavelength and $\eta $ ($\geqslant 1$) is
a index number. The only difference among the superconductors is the $%
\lambda _{e}$ of the paired electrons. Any quantum confinement of electrons
will reduce the $\lambda _{e}$ and help to form the paired electrons,
thereby, increase the superconducting temperature $T_{c}$.

In our opinion, it is the electronic helical motion which causes the pairing
of electrons and leads to superconductivity. The BCS theory \cite{bardeen}
of electron-phonon interaction is likely only to be a artificial mechanism
of superconductivity and the true mechanism may still be the simplest
electromagnetic interaction between the pairing electrons. For any stable
atoms of nature, the closed-shell electrons are naturally in pairing and the
corresponding atoms are the natural one-dimensional superconductors. For
traditional superconductors, pairing electrons can move in 3D space and they
are more freedom with a longer de Broglie wavelength $\lambda _{e}$, then,
the corresponding $T_{c}$ are lower. For the high-temperature
superconductors, the paired electrons are confined in a 2D plane orbit, this
confinement will greatly decrease $\lambda _{e}$ of the electrons,
consequently, increase the $T_{c}$.

Is it possible to increase the superconducting transformation temperature?
With our mechanism of superconductivity, the question turn out to be: how to
better confine the movement of paired electrons in the materials? High
pressure had been proved as a useful way of increasing the $T_{c}$ \cite%
{chu,shimizu,struzhkin,deemyad}, but it is not the final solution. Maybe
some artificial structures with fractional-dimension paired electrons orbits
are good choices \cite{bourgoin,gao,bao}.

\subsection{Mass of electron}

\begin{table*}[tp]
\caption{The relationship between the quantum confinement effect and the
mass of electron. }
\label{table2}\centering%
\begin{tabular}{cc|cc}
\hline
\multicolumn{2}{c|}{The less confinement systems} & \multicolumn{2}{|c}{The
more confinement systems} \\ \hline
Systems & Electron mass & Systems & Electron mass \\ \hline
$3D$ electronic orbit & $m_{e}\downarrow $ & $2D$ electronic orbit & $%
m_{e}\uparrow $ \\
$2D$ electronic orbit & $m_{e}\downarrow $ & $1D$ electronic orbit & $%
m_{e}\uparrow $ \\
outer-shell electron & $m_{e}\downarrow $ & inner-shell electron & $%
m_{e}\uparrow $ \\
wide-band electron & $m_{e}\downarrow $ & narrow-band electron & $%
m_{e}\uparrow $ \\
extended electronic state & $m_{e}\downarrow $ & localized electronic state
& $m_{e}\uparrow $ \\
increasing temperature & $m_{e}\downarrow $ & decreasing temperature & $%
m_{e}\uparrow $ \\
decreasing pressure & $m_{e}\downarrow $ & increasing pressure & $%
m_{e}\uparrow $ \\
low-speed helical movement & $m_{e}\downarrow $ & high-speed helical movement
& $m_{e}\uparrow $ \\
isotope containing more neutron & $m_{e}\downarrow $ & isotope containing
less neutron & $m_{e}\uparrow $ \\ \hline
\end{tabular}%
\end{table*}
In physics, the mass of the electron is associated with the maximum and most
complicated physical phenomena. Though we are much better prepared today
with copious amounts of definitions of the electron mass, for example, the
inertia and gravitational mass in Newtonian mechanics, relativistic mass in
Einstein theory, effective mass in semiconductor physics, electromagnetic
mass in electromagnetic theory and heavy fermion in condensed matter
physics. We think that a deeper physical understanding of this concept
(mass) is still not available in known physics. It seems difficult to
understand what is the mass of the electron and why the value of the
electron mass is not the intrinsic physical quantity.

In 1905, Einstein presented his famous formula, $E=mc^{2}$, known as the
energy-mass relation. In this paper, we have found another energy-mass
relation for the special particle of electron. Let us recall Eq. \ref%
{energy0}, what it says is that the mass of electron is associated with an
amount of magnetic energy. The most important conclusion related to our
energy-mass relation is, that a stronger quantum confinement of electron
(reducing the wavelength $\lambda _{e}$) will directly result in the
increment of the mass of the electron. Here we assume all known concepts of
electron mass would have the same physical reason. In order to have a deeper
understanding of the electron mass, we construct a table and make a
comparison between the electron mass of the less confinement systems and
that of the more confinement ones.

From Table. \ref{table2} one can see that all relationships presented are in
agreement with the known theoretical and experimental results. Therefore,
the so called quantum confinement effect is a procedure that leads to the
increment of the mass of electron, and an enhancement of the electron to
choose a more stable state. Our results indicate that the stability of the
electron can be measured by the mass of the electron. The more confinement
applied to a electron, the \textquotedblleft heavier\textquotedblright\ and
more stable it seems to be. To enhance comprehension of the correlation
between electron mass and its stability, it can be useful to consider some
specific examples. Special relativity says that a rapidly moving object will
have a heavier mass than the same object at a relatively lower speed (or at
rest). Our idea of this phenomenon is that the moving object has a tendency
to be more stable (because of the heavier mass). In fact, it is a common
knowledge that, like a bullet, high-speed helical movement can greatly
increase the stability of the bullet, and hence increase the mass of the
bullet. Next, let's consider the isotopic effect (the dependence of $T_{c}$
on the square root of isotopic mass), which was interpreted as experimental
evidence of the BCS theory of electron-phonon mediated. Physically, if the
atomic nucleus has more neutrons, in one hand the nucleus will become less
stable and in another hand it will weaken the interaction between nucleus
and electron. These will directly cause the instability of electrons,
consequently, decreasing both the mass of the pairing electrons and the
superconductivity temperature $T_{c}$.

In the above discussion, we have concluded that the mass of electron is the
effect of quantum confinement: a stronger quantum confinement of electron
will lead to the increment of the electron's mass, which is in good
agreement with the experimental results. In the next subsection, we will
show that the spin is no long as an intrinsic characteristic of electron and
the spin itself is also a quantum confinement induced effect.

\subsection{Spin, zero point energy, hardness property of electron and a
comparison of different electrons}

As well known, the rules for spin came from playing with experimental data.
So far the rules worked but remained mysterious. This mystery has inspired
vast theoretical and experimental activity to analyze and understand the
spin structure of the electron. But, in the past several decades scientists
have achieved little success in this challenging problem. Historically,
Uhlenbeck and Goudsmit \cite{uhlenbeck} (and separately Kronig) initially
proposed a physical picture for the electron as a spinning charged sphere of
radius equal to the so-called classical electron radius. According to this
model, the resulting speed of the surface of the electron would be greater
than the velocity of light. Up to today, it seems the classical image of a
body rotating about an axis is totally inadequate to describe the peculiar
geometrical properties of intrinsic spin. In our opinion, it is still
significative to ask: Where is the electron spin coming from? Do we really
have to accept that the spin is an intrinsic property of the electron with
no classical explanation for it whatsoever?
\begin{figure}[tbp]
\centering
\includegraphics[height=3.4in,angle=-90]{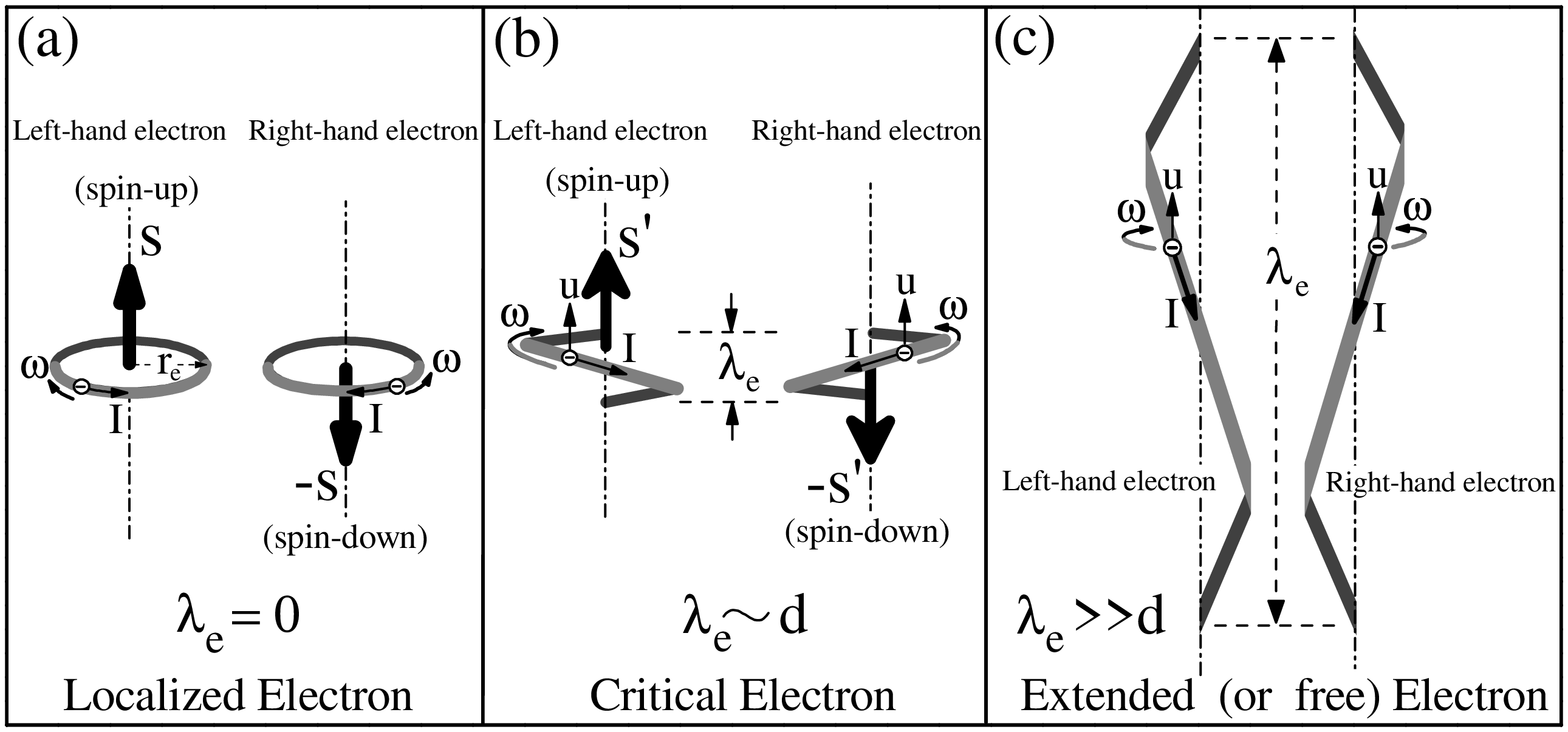}
\caption{ The relation between the electron's de Broglie wavelengths $%
\protect\lambda _{e}$ and the concept of electron spin $s$, where $d$ is the
atomic scale. (a) For a localized electron (or quantized \textquotedblleft
inducton\textquotedblright\ ), the spin $s$ can be well defined. (b) For a
critical electron, a quasi- spin $s^{\prime}$ can be approximately defined
and the corresponding electron is quantum electron. (c) For a free electron,
the spin cannot be defined and it will behave like a classical electron.}
\label{fig8}
\end{figure}

It's important to note that electrons have been physically divided into two
classes: the classical electrons (spin-independent), and the quantum
electrons (spin-dependent). Let's first recall the classical free electron
model (Fermi gases), which was good for explaining the Fermi energy, Fermi
surface, thermal property, specific heat, Hall effect and electrical
conduction of metals. For the quantum mechanical cases, a large variety of
spin-dependent transport effects appears in different regimes of condensed
matter physics: the Josephson junction \cite{josephson}, giant
magnetoresistance \cite{baibich}, fractional Hall effect \cite{tsui} and
spin Hall effect \cite{kato}, \textit{etc}. With the consummation of
sciences nowadays, it is not difficult to find that the so-called
spin-dependent phenomena can only be observed at very small scales (atomic
scales) and under some extreme conditions, such as ultra-low temperatures,
ultra-high pressures, ultra-high density, extremely magnetic fields. It is
quite strange that the spin-dependent behaviors are usually not seen at
larger scales. These evidently condition-dependent spin-related physical
properties inspire us to raise the following question: Is the artificial
concept of electron spin indeed \textquotedblleft
intrinsic\textquotedblright ? According to our logic it is without any doubt
that \textquotedblleft intrinsic spin\textquotedblright\ means the
properties of electron spin are condition-independent. Of course, an
electric charge is an intrinsic property of electrons, but spin should be
excluded. The electron spin, once thought to be intrinsic, now has an
extrinsic nature. We think it is about time for us to remove the shadow of
mystery cast upon the electron spin presented by the quantum theory. In the
following, we will clarify the origin of what physicists call
\textquotedblleft intrinsic spin\textquotedblright\ by the scheme of the
helical moving electrons.

In this paper, it has been suggested that a stronger quantum confinement on
electron will greatly reduce the electronic de Broglie wavelength $\lambda
_{e}$ and at the same time can help the electron toward a more localized
state (or stable state). Supposing $r_{e}\neq 0$, we show in Fig. \ref{fig8}
three kinds of electron (or inducton) of different de Broglie wavelengths.
Fig. \ref{fig8}(a) shows a localized electron that may only exist at the
lowest possible temperature (absolute zero). It is clear that, under this
extreme condition, the electron will be experienced the strongest quantum
confinement. As a result, the electron's de Broglie wavelength $\lambda _{e}$
and axial velocity $u$ are both equal to zero. In this special case, the
corresponding electrons can be considered as a localized ring current of
left-hand or right-hand according to the original chirality of the electron
of helical movement. As shown in Fig. \ref{fig8}(a), the ring can be
characterized by a small magnetic moment associated with a quantum number $s$
which, we think, is the equivalent parameter of the so-called intrinsic
electron spin defined by quantum theory. Evidently, the quantum number
(spin) $s$ can be either positive (spin-up) or negative (spin-down)
depending on the chirality of the electron used to define it. In addition,
we may note from the diagram of Fig. \ref{fig8}(a) that each ring contains a
certain amount electromagnetic energy $E_{s},$ which is likely the Zero
Point vibrational Energy (ZPE) presented in quantum mechanics. The ZPE
resulted from principles of quantum mechanics is so enormous that many
physicists have questioned: Is the zero point energy real? Supposing that
the concept of ZPE is physically genuine, what we are mostly interested in
whether or not our mechanism can provide a constructive interpretation of
the ZPE.

According to our theory, when a substance is cooled to absolute zero, inside
the substance the electron's de Broglie wavelength satisfies $\lambda
_{e}=0. $ Since the inductance $L_{e}$ ($r_{e}\neq 0$) of the current ring
of Fig. \ref{fig8}(a) always remains to be a finite quantity, then with the
Eq. \ref{inducton} we have $m_{e}\propto \lambda _{e}^{-2}$, consequently
the mass of electron $m_{e}\rightarrow \infty $ (as $\lambda _{e}\rightarrow
0).$ This suggests, according to Einstein's mass-energy relation, that an
absolute zero electron (or ring) contains an enormous amount of untapped
electromagnetic energy (infinite) known as zero point energy or ZPE. Though
our interpretation of ZPE seems consistent with that of quantum theory, but
the conclusion of infinite electron's mass implies that it is impossible to
have an absolutely ZPE condition.

In physics, there are few theories which claim to have universal
significance on all scales. It has been proved that the quantum mechanics is
valid for describing the bizarre rules of electron and light only at atomic
scales. Therefore, it becomes the first necessary to define a criterion by
which we might distinguish between classical systems and quantum systems.
After serious consideration, we think that the comparing the de Broglie
wavelength to the size of the object is a appropriate candidate of the
criterion. Figure \ref{fig8}(b) shows a traditional quantum system where the
electron's de Broglie wavelength ($\lambda _{e})$ has the atomic scale ($d$%
). We may note from the diagram that the quantized \textquotedblleft
inducton\textquotedblright\ can be uniquely presented by a quasi- spin
quantum number $s^{\prime},$ which averagely can also be either positive
(spin-up) or negative (spin-down) according to the chirality of the
electron. If we increase the ratio between $\lambda _{e}$ and $d,$ then the
both wave and spin characteristic of electron will gradually become fuzzy.
When $\lambda _{e}\gg d$, the studied system will be a real classical system
and the state of corresponding electron is much easy to be disturbed by the
extraneous factors. As a result, it is impossible to define both the
magnetic moment and the spin of the electron, and that is shown in Fig. \ref%
{fig8}(c).

Furthermore, we may note a particular case from the diagram of Figure \ref%
{fig8} with the radius of the helical orbit $r_{e}=0.$ For this
case, the so-called intrinsic electron's spin (strictly up and
down) can be defined for all three situations of the figure unless
$\lambda _{e}\rightarrow \infty .$ This implies that the quantum
mechanism is nothing more than a limit ($r_{e}\rightarrow 0)$ of
the $LC$ mechanism. A comparison of different electrons from the
$LC$ mechanism's point of view is given by: (i) Traditional
classical electron: $r_{e}=0$, and $\lambda _{e}\rightarrow \infty
;$ (ii) Quantum mechanism electron: $r_{e}=0$, and $\lambda _{e}$
has a finite value; and (iii) $LC$ mechanism electron: both
$r_{e}$ and $\lambda _{e}$ are finite.

Obviously the electron, described by quantum mechanism (ii), seems
physically inconsistent. As can be seen, on the one hand $r_{e}=0$ indicates
the vanishing of electron's wave property, on the other hand the finite $%
\lambda _{e}$ implies the existing electron's wave property. Quantum
mechanism overcame this difficulty by factitiously introducing some
uncertain and mysterious explanation of the microscopic particles.

Our results above reveal an essential connection between electron spin and
the intrinsic helical movement of electron and indicate that the spin itself
is the effect of quantum confinement, as well as provide new insights into
quantum theory. Furthermore, with these discussions it can be fairly easy to
interpret the hardness property of electron. Based on the well-known x-ray
techniques, the soft (low-energy or long wavelength) and hard x-rays
(high-energy or short wavelength) are defined. A similar situation exists
for the hardness property of electron. Electrons are either hard or soft
(only possibilities) depending upon the de Broglie wavelength of the
electrons. A harder electron has a shorter wavelength (or a heavier mass),
which implies that it contains more energy. Hence, the harder electron can
be regarded as a high-energy electron. Likewise, the softer electron (longer
wavelength or lighter mass) is a low-energy electron. This leads immediately
to a very important conclusion of hardness property of electron: the quantum
electron is harder and heavier than the classical electron. Because of the
high stability of the harder electron, obviously, the harder electron will
have both longer lifetime and longer coherent length. Otherwise, it has been
proved much more difficult to remove the inner electron than the outer
electron, our interpretation to this natural fact is: because the inner
electron is harder and at the same time heavier than the outer electron of
the same atom, therefore, the inner electron is harder to be removed.

\subsection{Magnetic monopole}

Now we will discuss the magnetic monopole. In 1931 Dirac \cite{dirac1}
introduced a magnetic monopole into the quantum mechanics and found a
quantization relation between an electric charge $e$ and magnetic charge $g$%
, $eg=\frac{n}{2}\hbar c,$ where $\hbar $ is the Plank's constant divided by
$2\pi $ and $n$ is an integer. Since then, numerous attempts of experimental
search for these magnetic monopoles at accelerators and in cosmic rays have
been done. Though, recently a multinational research group claimed that the
magnetic monopole can appear in the crystal momentum space \cite{fang}.
There has not yet been any firm evidence for its existence in real space.
Theoretically, 't Hooft \cite{hooft} pointed out that a unified gauge theory
in which electromagnetism is embedded in a semisimple gauge group would
predict the existence of the magnetic monopole as a soliton with spontaneous
symmetry breaking, Wu and Yang first described magnetic monopoles in terms
of a principal of fiber bundle \cite{wu}, Seiberg and Witten developed the
famous magnetic monopole equations \cite{seiberg}.

Here we are mostly concerned about the reason: Why have no magnetic
monopoles been detected after it had been hypothesized for 75 years? There
is a well-known reason: the magnetic monopoles are extremely heavy ($\sim $%
1016GeV) and well beyond the capabilities of any reasonable
particle accelerator to create. We don't think the mass of
magnetic monopoles can be a reasonable reason, personally, we are
more incline to think magnetic monopoles aren't naturally real. We
have an immature idea: Is the concept of magnetic monopole only a
well-known particle of different state? In fact, Seiberg-Witten
proved that there has an equivalent dual description through which
electron and magnetic monopole are interconvertible. In the
following discussion, it is shown that the magnetic monopole is,
in fact, a handed electron at absolute zero-temperature and
related to the \textquotedblleft spin\textquotedblright of
electron.

The huge mass of magnetic monopoles means that they are going to be pretty
slow (or more localized). Let us turn our attention to Fig. \ref{fig8}(a) of
tremendous electron's mass, when $r_{e}\rightarrow 0,$ the electron's
circular orbit will gradually disappear thus the handed electron will
degenerate into a structureless point charge of quantum theory. Eventually,
the so-called \textquotedblleft spin-up\textquotedblright\ and
\textquotedblleft spin-down\textquotedblright\ heavy electrons will turn out
to be the magic Dirac's magnetic monopoles with north (\textquotedblleft $%
N\textquotedblright $) and south (\textquotedblleft
$S\textquotedblright $) magnetic poles, respectively. Note the
case of $r_{e}=0$ is unallowed in our theory, hence, the concept
of magnetic monopoles is physically unreal. Now, I am confident
that any attempts for searching magnetic monopoles will be proved
completely\ valueless and furitless. Suppose $r_{e}=0$ (quantum
mechanism approximation), based on the above discussions, it is
clearly shown that the electron spin and magnetic monopole of
quantum mechanism are just two different expressions for one and
closely related thing of the mixed mechanism.

\section{Concluding Remarks}

In conclusion, we have found a process of perfect transformation of two
forms of energy (kinetic and field energy) inside the hydrogen atom and the
conservation of energy in the system. Then, we have shown that the helical
moving electron can be regarded as a inductive particle (\textquotedblleft
inducton\textquotedblright ) while atom as a microscopic $LC$ oscillator
then the indeterministic quantum phenomena can be well explained by the
deterministic classical theory. In particular, we have show that a pairing
Pauli electron can move closely and steadily in a DNA-like double helical
electron orbit. Moreover, we have pointed out that the mass of electron, the
ZPE and what has been called the intrinsic \textquotedblleft electron
spin\textquotedblright\ are all really the quantum confinement effects of
the intrinsic chirality of the electron of helical motion. For
superconductivity, we should be able to confine electron pairs to the low
dimensional systems and produce a higher temperature superconductor.
Finally, we show analytically that the magnetic monopole is, in fact, a
special handed electron ($r_{e}=0$) at absolute zero-temperature with a de
Broglie wavelength $\lambda _{e}=0.$ This result indicates that any attempts
to search for magnetic monopole in real space will be proved to be in vain.
In addition, we have pointed out that the quantum mechanism's concepts of
electron's spin and magnetic monopole are just two different expressions for
one possible physical phenomenon.

We have shown that the quantum mechanism is nothing but an approximate
theory (with the radius of the helical orbit $r_{e}\rightarrow 0)$ of the $%
LC $/wave-particle duality mixed mechanism. Our mixed mechanics
force us to rethink the nature and the nature of physical world.
We believe all elementary particles, similar to photon and
electron, are only some different types of energy representation.
Though, the standard quantum mechanics nature is intrinsically
probabilistic, permitting only predictions about probabilities of
the occurrence of an event. Nevertheless, one century after its
birth, it still presents many unclarified issues at its very
foundations. Starting from an Einstein's work \cite{epr}, many
attempts have been devoted to build a deterministic theory
reproducing all the results of quantum mechanics. The latter
include the de Broglie-Bohm's hidden variable theory, the most
successful attempt in this sense \cite{bohm}. Recently, a first
experimental test of de Broglie-Bohm theory against standard
quantum mechanics was reported \cite{brida}. In our study, it has
been shown definitely that the electron follows a perfectly
defined trajectory in its motion, which confirms the de
Broglie-Bohm's prediction. Also in our work, it is found that the
known wave-particle duality can be best manifested by showing that
the wave motion associated with a electron is just the phenomenon
of its complex helical motion in real space. Therefore, the
wave-particle duality should lie at the heart of the quantum
universe. We are now more and more convinced that the universe was
built in the simplest manner and all things in it are unique and
definitive. As Albel Einstein one said, \textquotedblleft God does
not play dice with the universe\textquotedblright . Of course, a
more clear understanding of microscopic world is still of the
greatest challenge.

\bigskip \textbf{Acknowledgement:} The author would like to thank Ron
Bourgoin for valuable discussions. This work was supported by the grants
from National Nature Science Foundation of China (90201039, 10274029).


\begin{thebibliography}{99}
\bibitem{planck} M. Planck, Ann. Phys. \textbf{1}, 69 (1900).

\bibitem{einstein} A. Einstein, Ann. Phys. \textbf{17}, 132 (1905).

\bibitem{bohr} N. Bohr, Phil. Mag. \textbf{26}, 576 (1913).

\bibitem{dirac} P. A. M. Dirac, \textit{The principles of Quantum Mechanics}%
. Oxford U. Press (1958).

\bibitem{stern} O. Stern, Z. Phys. \textbf{2}, 49 (1920).

\bibitem{schrodinger} E. Schrodinger, Ann. Phys. \textbf{79}, 361 (1923).

\bibitem{broglie} L. de Broglie, Phil. Mag. \textbf{47}, 446 (1924).

\bibitem{pauli} W. Pauli, Z. Phys. \textbf{31}, 373 (1924).

\bibitem{uhlenbeck} G. E. Uhlenbeck and S. Goudsmit, Naturwissenschaften
\textbf{47}, 953 (1925).

\bibitem{heisenberg} W. Heisenberg Z. Phys. \textbf{43}, 172 (1927).

\bibitem{davisson} C. Davisson, L. H. Germer, Phys. Rev. \textbf{30}, 707
(1927).

\bibitem{thomson} G. P. Thomson, Proc. Roy. Soc. \textbf{117}, 600 (1928).

\bibitem{lamb} W. E. Lamb, R. C. Retherford, Phys. Rev. \textbf{72}, 241
(1947).

\bibitem{carreyre} T. Carreyre, et al., Phys. Rev. C \textbf{62}, (2000).

\bibitem{kishimoto} S. Kishimoto, et al., Phys. Rev. Lett. \textbf{85}, 1831
(2000).

\bibitem{zeeman} P. Zeeman, Phil. Mag. \textbf{43}, 226 (1897).

\bibitem{watson} J. D. Watson, F. H. C. Crick, Nature \textbf{171}, 737
(1953).

\bibitem{morris} M. Morris, K. Uchida and T. Do, Nature \textbf{440}, 308
(2006).

\bibitem{chu} C. W. Chu, P. H. Hor, R .L. Meng, L. Gao, Z. J. Huang and J.
Q. Wang, Phys. Rev. Lett. \textbf{58}, 405 (1987).

\bibitem{shimizu} K. Shimizu, H. Ishikawa, D. Takao, T. Yagi,m and K. Amaya,
Nature \ \textbf{419}, 597 (2002).

\bibitem{struzhkin} V. V. Struzhkin, M. I. Eremets, W. Gan, H-K. Mao, and R.
J. Hemley, Science \textbf{298}, 1213 (2002).

\bibitem{deemyad} S. Deemyad and J. S. Schilling, Phys. Rev. Lett. \textbf{91%
}, 167001 (2003).

\bibitem{bourgoin} R. Bourgoin, http://neasia.nikkeibp.com/neasia/002073,
(2005).

\bibitem{gao} Y. Gao, et al., Science \textbf{306, }1915 (2004).

\bibitem{bao} X. Y. Bao et al., Phys. Rev. Lett. \textbf{95}, 247005 (2005).

\bibitem{bardeen} J. Bardeen, L. N. Cooper and J. R. Schrieffer, Phys. Rev.
\textbf{108}, 1175 (1957).

\bibitem{josephson} B. D. Josephson, Phys. Lett. \textbf{1}, 251 (1962).

\bibitem{baibich} M. N. Baibich, J. M. Broto and A. Fert, Phys. Rev. Lett.
\textbf{61,} 2472 (1988).

\bibitem{tsui} D. C. Tsui, H. L. Stormer, and A. C. Gossard, Phys. Rev.
Lett. \textbf{48}, 1559 (1982).

\bibitem{kato} Y. K. Kato, R. C. Myers, A. C. Gossard, and D. D. Awschalom,
Science \textbf{306}, 1910 (2004).

\bibitem{dirac1} P. A. M. Dirac, Proc. Roy. Soc. \textbf{A133}, 60 (1931).

\bibitem{fang} Z. Fang et al., Science \textbf{302}, 92 (2003).

\bibitem{hooft} G. 't Hooft, Nucl. Phys. \textbf{B79}, 276 (1974).

\bibitem{wu} T. T. Wu and C. N. Yang, Phys. Rev. D \textbf{12}, 384 (1975).

\bibitem{seiberg} N. Seiberg and E. Witten, Nucl. Phys. B \textbf{431}, 484
(1994).

\bibitem{epr} A. Einstein, B. Podolsky and N. Rosen, Phys. Rev. \textbf{47},
777 (1935).

\bibitem{bohm} D. Bohm, Phys. Rev. \textbf{85} 166 (1952); D. Bohm, Phys.
Rev. \textbf{85} 180 (1952).

\bibitem{brida} G. Brida, et al., J. Phys. B \textbf{35}, 4751 (2002).
\end{thebibliography}
\end{document}